\documentclass[cameraready]{Interspeech}
\usepackage{xcolor}

\usepackage{graphicx} % \resizebox 
\usepackage{float} % [H] option
\usepackage{placeins} % \FloatBarrier
\usepackage{stfloats} % two-column mode - table top fix
\usepackage{subcaption}

\title{Designed Vocalizations Dataset: Sound-Designed Human and Animal Voices for Non-human Voice Conversion}

\author[affiliation={1}]{Seolhee}{Lee}
\author[affiliation={1}]{Minsu}{Kang}
\author[affiliation={1}]{Yangsun}{Lee}
\author[affiliation={1}]{Woosun}{Min}
\author[affiliation={1}]{Choonghyeon}{Lee}
\author[affiliation={1,2}]{Namhyun}{Cho}

\address{
    $^1$ NC AI Co., Ltd, Republic of Korea \\
    $^2$ Sogang University, Republic of Korea
}

\email{\{seolhee, mskang, yyslee0301, choonghyeon, cnh2769\}@ncsoft.com, woosun1108@naver.com}


\usepackage{comment}

\begin{document}

\maketitle

% the abstract here must exactly match the abstract entered into the paper submission system
\begin{abstract}    

    % 1000 characters. ASCII characters only. No citations.
    Advances in AI-based voice conversion have enabled a wide range of media applications, including films, audiobooks, and games. However, most research and public benchmarks still focus on natural human speech, leaving designed vocalizations, such as monster growls and robotic voices, underexplored, partly due to the lack of publicly available resources. To address this gap, we introduce the Designed Vocalizations Dataset, constructed by curating diverse raw vocal sources, including speech and animal vocalizations, and applying professional vocal effects processing to produce corresponding effect-modified variants. We further provide a standardized test set with explicit seen/unseen splits over source timbre groups and preset styles to assess generalization under controlled conditions. Finally, we report baseline benchmark results to support reproducible evaluation and future research. The dataset and demo samples are available online. \footnote{\url{https://ncai-official.github.io/speech/publications/designed-vocalizations-dataset/}}
\end{abstract}

\section{Introduction}
    The growth of interactive and creative media industries (games, film, animation, VR/AR, etc.) has driven increasing demand for diverse vocal sounds that enhance character expression and immersion. In particular, non-natural/non-human vocalizations such as monster roars, robotic voices, and stylized character utterances are essential for shaping a content's identity and narrative atmosphere. In production, these vocalizations are difficult to capture through recording alone. Sound designers typically rely on complex DSP chains, including distortion, spectral transformation, modulation, and multitrack mixing, which require iterative construction and careful tuning. Consequently, producing high-quality designed vocalizations remains a labor-intensive manual workflow, motivating the development of automated or assistive tools.

    Recent work has proposed human-to-non-human voice conversion (H2NH-VC) methods targeting non-natural, non-human timbres~\cite{Speak-like-Dog, samoye, humansgrowl, cartoonsing}. However, most of these efforts rely on internally curated datasets and evaluation resources, which are rarely released publicly. Data compositions and evaluation protocols also differ across studies, making fair comparison under matched conditions difficult. This lack of open resources contrasts with natural speech research, where dedicated public datasets~\cite{vctk, libritts, librilight, hifitts} and benchmarks have enabled reproducible evaluation and accelerated progress across TTS and voice conversion~\cite{tacotron2, fastspeech2, vits, yourtts, freevc, HierVST, Diff-HierVC, DDDM-VC}. To support systematic comparison and generalization tests in the non-human domain, publicly available datasets and standardized benchmarks are still needed. 

    To address this gap, we introduce the Designed Vocalizations Dataset, a public dataset of sound-designed vocalizations. The dataset covers linguistic sources and diverse non-linguistic vocalizations, including animal sounds, interjections, and vocal mimicry. In addition to raw recordings, it includes designed vocalizations used in real-world sound design. The designed vocalizations are produced by professional sound designers using effect-chain presets, spanning styles such as monster and creature vocalizations, robotic voices, low-register power voices, and metallic or abrasive voices. To support analysis and reproducibility, we provide preset-level effect summaries for the built-in presets, along with full effect-chain structures and parameter settings for the in-house-designed presets.

    The dataset provides a dedicated test set that supports both objective evaluation and listening tests. Each test item consists of a source and its preset-specific designed reference; the model generates an output that preserves the content of while matching the timbre style of the designed reference. Here, the source refers to the input vocalization to be converted, and the reference is the preset-specific designed version of the same source. The test set also includes explicit seen/unseen splits across preset styles and source timbre groups, enabling systematic analysis of generalization to unseen styles and unseen sources. Finally, we evaluate a representative voice conversion model~\cite{humansgrowl} on this benchmark and report baseline performance, providing a reference point for reproducible comparison in future work.

    The main contributions of this paper are as follows.    
    \begin{itemize}
        \item \textbf{Publicly available designed vocalization dataset.}
        We introduce the \textit{Designed Vocalizations Dataset}, a publicly available dataset containing designed vocalizations produced by professional sound designers. The dataset covers diverse linguistic and non-linguistic vocal sources, including speech, animal-like vocalizations, interjections, and vocal mimicry.

        \item \textbf{Standardized test set with (source-reference) pairs.}
        The test set provides (source, reference) pairs that enable evaluation of voice conversion models, where the model is assessed on how well it reflects the designed timbre (style) of the reference in the converted source.
        
        \item \textbf{Benchmark splits for generalization analysis.}
        Explicit seen/unseen splits across preset styles and source timbre groups enable systematic evaluation of generalization to unseen styles and unseen sources.
        
        \item \textbf{Baseline results on the proposed benchmark.}
        We report baseline results of a representative voice conversion model on the proposed benchmark to facilitate reproducible comparison in future work.
    \end{itemize}
    
\section{Designed Vocalizations Dataset}
    \begin{figure}[t]
      \centering
      \includegraphics[width=\linewidth]{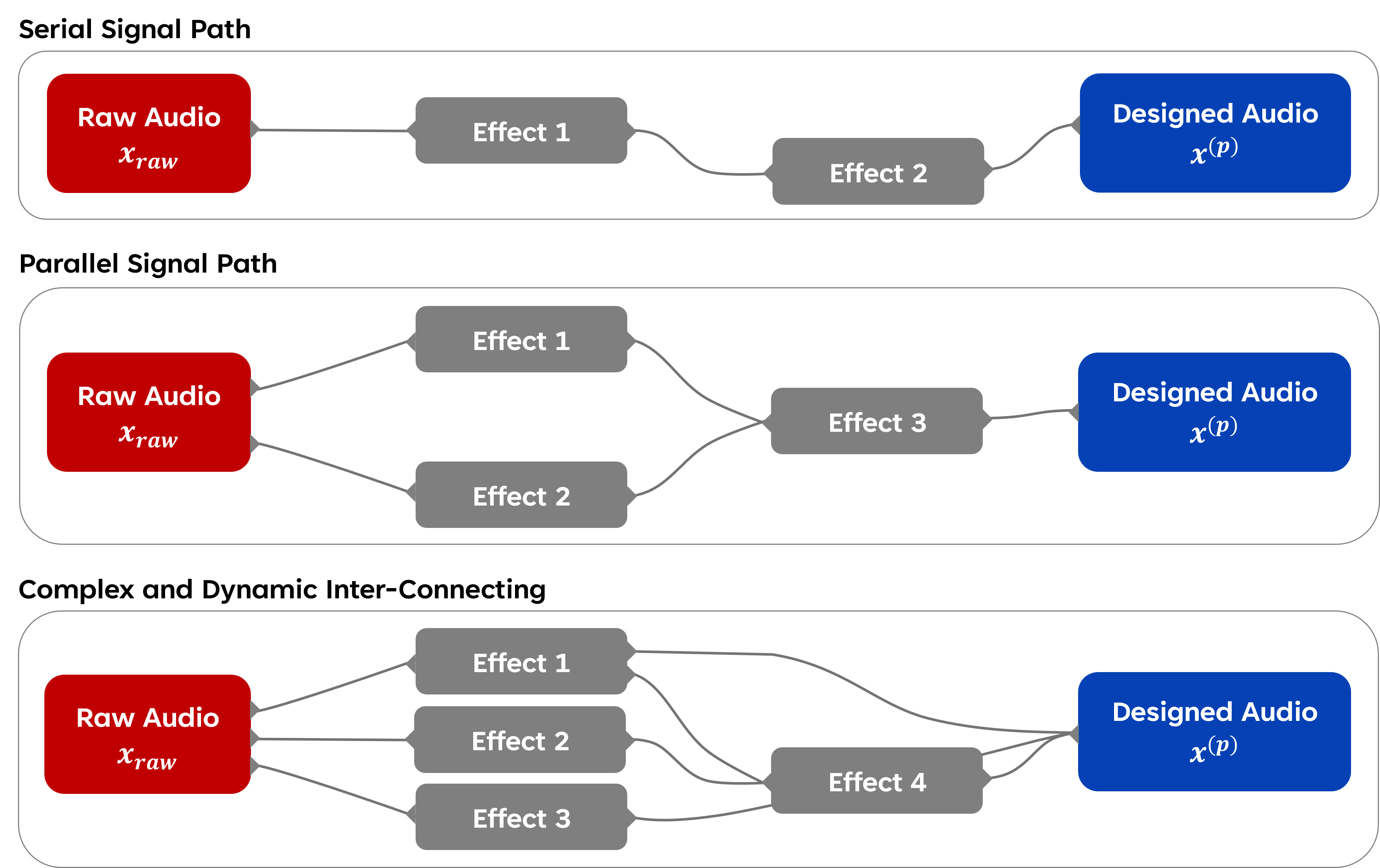}
      \caption{Representative audio effect chain structures for preset-specific DSP operators $G_p$. Each operator maps a raw vocal source $x_{\mathrm{raw}}$ to a designed vocalization $x^{(p)}$ using selected effect modules $\mathcal{M}_p$ connected in serial, parallel, or hybrid forms.}
      \label{fig:preset_diagram}
    \end{figure}

    \subsection{Description of Human-to-Nonhuman Voice Conversion}
        \label{sec:description-of-nhvc}
        Given a source vocalization $x_s$ and a reference vocalization $x_r$, a conversion model $F_{\theta}$ generates $\hat{y}=F_{\theta}(x_s,x_r)$, where $\hat{y}$ is expected to preserve the content and temporal structure of $x_s$ while reflecting the target timbre or style represented by $x_r$. In conventional voice conversion, both $x_s$ and $x_r$ are typically natural human speech samples. In the H2NH-VC task, $x_s$ is an unprocessed raw source vocalization, including speech, interjections, animal vocalizations, or vocal mimicry, whereas $x_r$ is a reference designed vocalization produced by applying sound-design processing to a raw vocal source.
        
        In the dataset construction process, we represent each designed vocalization as $x^{(p)} = G_p(x_{\mathrm{raw}})$, where $x_{\mathrm{raw}}$ is raw vocal audio, $p\in\mathcal{P}$ denotes a sound-design preset, and $G_p$ is the preset-specific DSP operator that maps the raw vocal audio to a designed vocalization. Each $G_p$ is built from selected effect modules $\mathcal{M}_p=\{m_{1}^{(p)},\ldots,m_{K_p}^{(p)}\}$, where $K_p$ is the number of modules used in preset $p$. The operator $G_p$ represents the complete effect chain, including the serial, parallel, or hybrid connections among these modules, as illustrated in Fig.~\ref{fig:preset_diagram}. In the test set, the reference sample $x_r$ is obtained by applying the preset-specific operator $G_p$ to the source vocalization $x_s$, i.e., $x_r=G_p(x_s)$, and specifies the target designed style to be reflected in the converted output.
        
    \subsection{Dataset Overview}
        
        The Designed Vocalizations Dataset is a publicly available dataset for research on timbre transfer and reproduction of sound design vocalizations. It consists of raw vocalizations $x_{\mathrm{raw}}$ and their designed variants $x^{(p)}=G_p(x_{\mathrm{raw}})$, generated by applying preset-specific sound-design processing to the raw sources. The raw sources include speech and a range of non-linguistic vocalizations, such as animal-like vocalizations, interjections, and vocal mimicry, covering the input distribution encountered in real-world content production. Designed vocalizations were generated using Dehumaniser~2~\cite{dehumaniser}\footnote{Dehumaniser~2 is a professional audio processing tool for creating non-human, monster-like, and stylized vocal effects using multiple audio effect modules.}. The preset set includes Dehumaniser~2 built-in presets and presets designed in-house. Some in-house-designed presets additionally incorporated post-processing in Cubase~\cite{cubase}.

        The dataset is organized into train and test splits. The train split provides raw and designed audio in a non-parallel format, supporting training setups that do not rely on paired data. The test split consists of (source, reference) pairs, where models are evaluated on their ability to generate outputs that match the designed timbre (style) of the reference while preserving the content of the source. Explicit seen/unseen splits across preset styles and source timbre groups are also provided, enabling evaluation under various generalization conditions.

    \subsection{Construction}
        \subsubsection{Source Audio Data Collection and Refinement}
        \label{raw-audio-collection}
        The source audio was categorized into linguistic and non-linguistic data. Linguistic samples were randomly selected from the VCTK dataset~\cite{vctk}, with 30 samples from each of 109 speakers, totaling 3,270 samples. Non-linguistic data included animal sounds, sounds mimicking animals and monsters, interjections, mumbles, and infant vocalizations sourced from Freesound~\cite{freesound} through keyword searches. Only samples under 5 seconds with a minimum sampling rate of 44.1 kHz were selected.
        
        Each audio file contained multi-tag metadata, and tag-based filtering was applied to exclude unsuitable samples, such as environmental sounds, background noise, music, and mixed-object recordings. Additionally, human reviewers manually removed irrelevant samples that were not captured by tag-based filtering. Automatic denoising was applied to eliminate noise, and volume normalization was performed to balance levels across the Freesound data.

        \subsubsection{Designed Vocalization Production}
            
            Designed vocalizations were produced by applying the preset-specific DSP operators $G_p$ defined in Section~\ref{sec:description-of-nhvc} to the curated raw audio. In practice, each $G_p$ corresponds to a complete sound-design chain rather than a single effect, combining multiple effect modules $m_k^{(p)}$.
            
            The processing was implemented using Dehumaniser~2~\cite{dehumaniser}. Specifically, 30 built-in presets and 17 in-house-designed presets were used to produce diverse designed vocalizations. The in-house-designed presets additionally included post-processing in Cubase~\cite{cubase}, such as reverb, compressor, limiter, distortion, and EQ. Fig.~\ref{fig:preset_diagram} illustrates representative effect-chain configurations, including serial, parallel, and hybrid structures. The primary effect modules corresponding to $m_{k}^{(p)}\in\mathcal{M}_{p}$ are described below.
            
            \begin{itemize} 
                \item \textbf{Delay Pitch Shifting}: Replays the input signal through a delay loop while shifting pitch for each loop, thereby creating a continuously varying pitch. Adjustable parameters include feedback, octave offset, and delay time, which control the number of repetitions and the degree of pitch alteration. 
                \item \textbf{Flanger/Chorus}: Uses modulated delay times to produce a flanger effect (robotic sound) with short delays and a chorus effect (vocal ensemble) with longer delays. Parameters include waveform, pitch depth, frequency, delay time, mix, feedback, and the number of voices, which control texture. 
                \item \textbf{Granular}: Divides input into small grains with varying pitch and texture, producing effects from whispery to rough. Parameters include grain pitch, size, density, and variation.
                \item \textbf{Noise Generator}: Distorts input to create harsh, aggressive sounds. Bandwidth and harmonic index adjustments enable distinctive effects such as robotic timbres.
                \item \textbf{Pitch Shifting}: Modifies pitch with precise octave, semitone, and cent controls.
                \item \textbf{Ring Modulator}: Multiplies input with an oscillator signal for classic ring modulation, adding robotic timbres. Controls include modulation depth, oscillator frequency, and waveform blending using a wheel, blend, and LFO. 
                \item \textbf{Spectral Shifting}: Alters pitch within a defined frequency range for unique spectral transformations. Thresholds and semitone transpositions control the range and degree of pitch change.
            \end{itemize}
        
    \subsection{Characteristics and Statistics}
        The dataset provides: (1) non-parallel training data consisting of raw and designed vocalizations, (2) aligned pairs of (source, reference) samples, and (3) metadata for the original Freesound~\cite{freesound} audio, including ID, filename, duration, and tags.\footnote{The released metadata includes preset-level effect summaries for built-in presets and full effect-chain structures with parameter settings for in-house-designed presets.} All audio files were standardized to 44.1 kHz, 16-bit WAV format, totaling 237,574 samples.

    \begin{table}[t]
      \caption{Overall data composition.}
      \label{tab:overall_data_composition}
      \setlength{\tabcolsep}{4pt}
      \begin{tabular}{@{} l l l l @{}}
        \toprule
        \textbf{Dataset} & \textbf{Presets} & \textbf{Raw Sources} & \textbf{Designed} \\
        \midrule
        Train &
        40 &
        \begin{tabular}[t]{@{}l@{}}5,654\\
          \quad Non-ling.: 2,384\\
          \quad Ling.: 3,270\end{tabular} &
        226,160 \\
        \midrule
        Test &
        \begin{tabular}[t]{@{}l@{}}47\\
          \quad Seen: 40\\
          \quad Unseen: 7\end{tabular} &
        \begin{tabular}[t]{@{}l@{}}120\\
          \quad Seen: 60\\
          \quad \quad Non-ling.: 30\\
          \quad \quad Ling.: 30 (VCTK)\\
          \quad Unseen: 60\\
          \quad \quad Non-ling.: 30\\
          \quad \quad Ling.: 30 (HifiTTS)\end{tabular} &
        5,640 \\
        \midrule
        Total & -- & 5,774 & 231,800 \\
        \bottomrule
      \end{tabular}
    \end{table}

    \subsubsection{Composition}

        The dataset is divided into training and test sets, with the overall composition summarized in Table~\ref{tab:overall_data_composition}. The training set consists of 40 preset conditions and 5,654 raw vocalizations. During training, raw and designed vocalizations are provided in a non-parallel setting. In total, the training set contains 226,160 designed vocalizations. The test set provides aligned (source, reference) pairs for quantitative evaluation and controlled listening tests. Here, source denotes the raw vocalization used as the conversion input, while reference is an audio sample that characterizes the target designed timbre. This setup enables the evaluation of timbre reproduction at the output level, independent of whether the model explicitly estimates the underlying effect-chain parameters. Specifically, each of the 120 test sources $x_s$ is processed with 47 presets $p$, producing preset-specific designed references $x_r=G_p(x_s)$. These presets include 40 observed during training (seen) and 7 held out from training (unseen), resulting in a total of $120 \times 47 = 5{,}640$ samples. The test inputs (source) are composed as follows:

        \begin{itemize} 
            \item \textbf{Source-seen}: 60 samples from 20 timbre types (10 linguistic, 10 non-linguistic) present in the training set. Linguistic samples are drawn from 10 VCTK~\cite{vctk} speakers.
            
            \item \textbf{Source-unseen}: 60 samples from 20 previously unseen timbre types (10 linguistic, 10 non-linguistic). Linguistic samples are drawn from 10 HiFiTTS~\cite{hifitts} speakers.
        \end{itemize}
        
        For non-linguistic inputs, the seen/unseen status was determined via tag-based category classification, followed by manual listening verification.
    
    \begin{figure}[t]
      \centering
      \includegraphics[width=\linewidth]{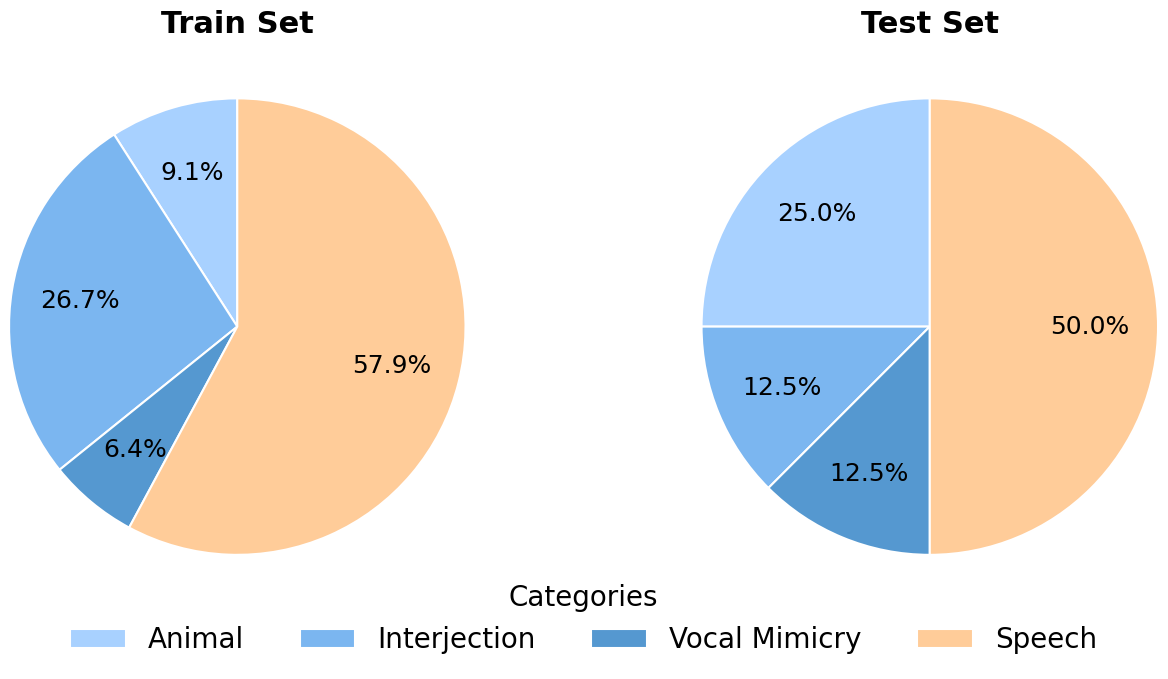}
      \caption{Source category distribution of train and test sets.}
      \label{fig:source_category_distribution}
    \end{figure}
    
    \subsubsection{Diversity of Raw Vocalizations}
        The raw audio consists of four categories: speech, interjections, animal sounds, and vocal mimicry. 
        The \textit{interjection} category includes vocalizations such as laughter, crying, and coughing, as well as mixed vocalizations containing partial speech and mumbling. 
        \textit{Vocal mimicry} is defined as human vocalizations that imitate monster or animal sounds and is distinguished from naturally occurring animal sounds. As shown in Fig.~\ref{fig:source_category_distribution}, the training data are dominated by speech (57.9\%), followed by interjections (26.7\%), animal sounds (9.1\%), and vocal mimicry (6.4\%). The test data are constructed to maintain representative coverage while ensuring diversity, consisting of speech (50.0\%), animal sounds (25.0\%), interjections (12.5\%), and vocal mimicry (12.5\%).
        
    \subsubsection{Diversity of Designed Vocalizations}
\begin{figure}[t]
        \centering
        \includegraphics[width=\linewidth]{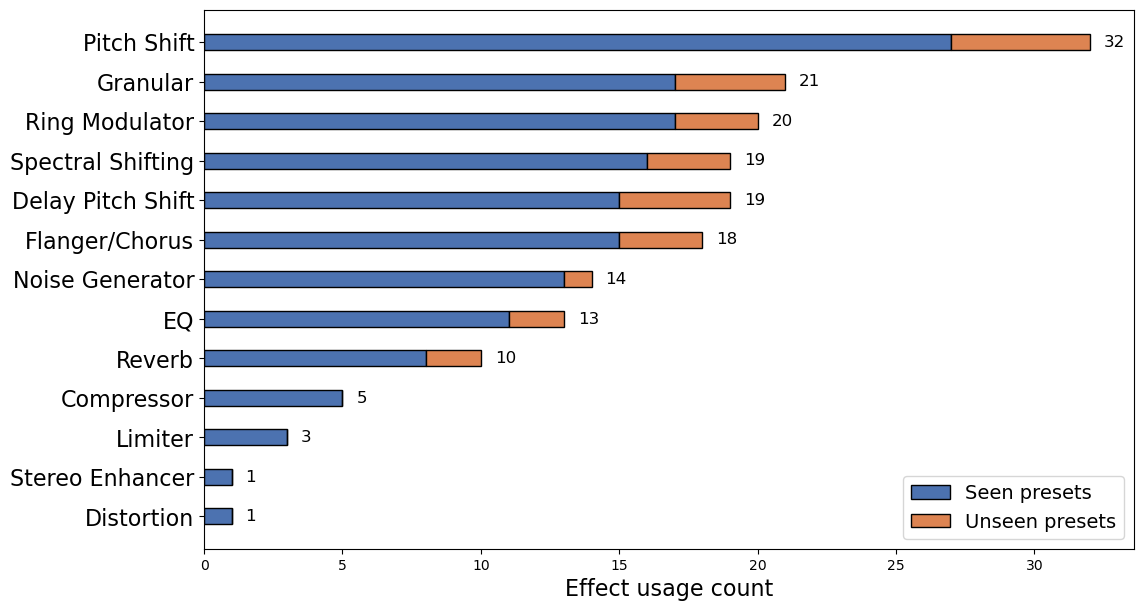}
        \caption{Effect usage distribution across seen and unseen preset groups. Top: seven Dehumaniser~\cite{dehumaniser} base effects; bottom: six additional effects in in-house-designed presets.}
        \label{fig:effect_distribution}
    \end{figure}

        The presets in Dehumaniser~2~\cite{dehumaniser} are organized into several categories representing different timbre styles, including robotic, creature-like, monster-like, speaker-like, and glitch-based sounds. Each preset produces distinct timbral characteristics through different combinations of effect chains and parameter settings. Presets were selected across these style categories, and in-house-designed presets were included alongside the built-in presets to expand the preset space.
        Built-in presets are composed of combinations of seven core effect modules provided by Dehumaniser~2~\cite{dehumaniser}, while in-house-designed presets incorporate six additional external effects not present in the built-in presets. Of the 47 presets in total, 40 presets (25 built-in and 15 in-house-designed) are used for training and treated as seen presets ($p\in\mathcal{P}_{\mathrm{seen}}$), while the remaining 7 presets (5 built-in and 2 in-house-designed) are held out as unseen presets ($p\in\mathcal{P}_{\mathrm{unseen}}$). Testing covers all 47 presets, including both seen and unseen preset styles.
        The training set was constructed to avoid the over-representation of any particular effect module, while the test set was designed so that each effect module is more evenly represented. The distribution of effect usage across all presets is shown in Fig.~\ref{fig:effect_distribution}.

\section{Experiment}
    
    % table
    \begin{table*}[t]
        \caption{Objective and subjective evaluation results across conversion scenarios.}
        \label{tab:eval_results_combined}
        \centering
        \setlength{\tabcolsep}{6pt}
        \begin{tabular}{lcccccc}
            \toprule
            & \multicolumn{5}{c}{\textbf{Objective}} & \multicolumn{1}{c}{\textbf{Subjective}} \\
            \cmidrule(lr){2-6}\cmidrule(lr){7-7}
            \textbf{Scenario}
            & \textbf{Cos. Sim. ($\uparrow$)}
            & \textbf{PCC-E ($\uparrow$)}
            & \textbf{RMSE-E ($\downarrow$)}
            & \textbf{CER ($\downarrow$)}
            & \textbf{WER ($\downarrow$)}
            & \textbf{MOS ($\uparrow$)} \\
            \midrule
            Seen-to-seen      & 0.667 & 0.985 & 0.047 & 3.79\% & 7.55\% & 3.81 \\
            Seen-to-unseen    & 0.648 & 0.984 & 0.049 & 3.59\% & 7.38\% & 3.66 \\
            Unseen-to-seen    & 0.643 & 0.982 & 0.047 & 1.89\% & 5.23\% & 3.66 \\
            Unseen-to-unseen  & 0.610 & 0.983 & 0.049 & 1.64\% & 4.65\% & 3.49 \\
            \bottomrule
        \end{tabular}
    \end{table*}
        
    This section discusses the evaluation of the quality and usability of the proposed dataset by assessing the performance of the voice conversion model and identifying directions for future research. The experiments assessed model performance on the four seen/unseen combinations of source $x_s$ and reference $x_r=x^{(p)}$ using the test set, following training on the training set. The evaluation measured the model's ability to convert voices using both seen preset styles and held-out preset styles.

    \subsection{Experimental Setup}
        We present benchmark performance using the recently proposed human-to-non-human voice conversion model (H2NH-VC)~\cite{humansgrowl} as a representative baseline for designed vocalization conversion.
        The H2NH-VC model retains the Conditional adversarial VAE (CVAE) architecture but incorporates adjusted preprocessing to capture finer acoustic details. Specifically, the Short-Time Fourier Transform (STFT) parameters were modified, setting the frame/window length and hop length to 20 ms and 5 ms, respectively, to achieve a finer temporal resolution. Style embedding was applied exclusively to the prior network and flow module to enhance timbre conversion. Training was conducted using the multi-resolution Mel-STFT loss~\cite{DAC} covering a wide frequency range from 0 to 22.05 kHz.

    \subsection{Evaluation Metrics}
         Model performance was assessed using objective and subjective metrics. Objective evaluation included: (1) cosine similarity using time-averaged BEATs~\cite{BEATs} embeddings from SALMONN~\cite{SALMONN} to measure timbre similarity between the converted output $\hat{y}$ and the reference audio $x_r$, (2) Pearson correlation coefficient and root mean squared error for energy (PCC-E and RMSE-E) between $\hat{y}$ and the source $x_s$ to evaluate energy prosody preservation, and (3) character error rate (CER) and word error rate (WER) using Whisper~\cite{Whisper} to quantify pronunciation clarity and recognition accuracy. A subjective listening test was conducted using a 5-point Mean Opinion Score (MOS) with 8 participants evaluating 40 samples. Participants were asked to assess whether the converted outputs reached a perceptual quality suitable for interactive media applications.
         
    \subsection{Results and Analysis}
        Table~\ref{tab:eval_results_combined} reports benchmark results under four scenarios (seen--seen, seen--unseen, unseen--seen, unseen--unseen) defined by the seen/unseen combinations of the source and the preset (reference). The MOS score decreases from 3.81 (seen--seen) to 3.49 (unseen--unseen), with the cross settings both at 3.66, reflecting a gradual drop in perceived quality under more challenging conditions. For timbre similarity, cosine similarity (Cos. Sim.) is highest in seen--seen (0.667) and lowest in unseen--unseen (0.610), with the two cross settings in between (seen--unseen: 0.648; unseen--seen: 0.643), indicating increased difficulty when unseen factors are introduced. In contrast, the energy-prosody preservation metrics show little variation across scenarios: PCC-E stays within 0.982--0.985 and RMSE-E within 0.047--0.049, suggesting limited sensitivity to the seen/unseen split in this benchmark. Finally, the ASR-based intelligibility metrics (CER/WER) are lower when the source is unseen (unseen--seen and unseen--unseen; CER: 1.89\%/1.64\%, WER: 5.23\%/4.65\%). We speculate that in these more challenging settings, weaker conversion by $F_{\theta}$ may leave the output $\hat{y}$ closer to the original source characteristics, which can be easier for ASR to recognize than fully designed timbres.

\section{Conclusion}

    % [기존]
    %This study addresses the lack of public datasets and benchmarks for designed vocalizations by introducing a dataset for training and evaluating designed vocalizations. The dataset includes both raw audio and preset-specific designed variants generated through sound-design operators, incorporating diverse linguistic and non-linguistic sources. We report baseline benchmark results to facilitate fair comparison across systems. Future work may expand the dataset to cover a broader range of effects and encourage model improvements for effect-aware conversion. We hope that our proposed dataset and benchmark support the research community in advancing vocalization synthesis beyond speech, including diverse non-speech and designed vocal sounds.

    % [변경]
    This study introduces the Designed Vocalizations Dataset, a public dataset and benchmark for training and evaluating designed vocalization conversion models. The dataset includes raw audio and preset-specific designed variants generated through sound-design operators, covering diverse linguistic and non-linguistic sources. We report baseline results to facilitate fair comparison across systems. Future work may expand the dataset to cover broader effect types and encourage effect-aware model improvements. We hope this dataset and benchmark support research on vocalization synthesis beyond speech, including diverse non-speech and designed vocal sounds.

\newpage

\section{Acknowledgments}

This work was partly supported by the Institute of Information \& Communications Technology Planning \& Evaluation (IITP) grant funded by the Korea government (MSIT) (No. RS-2025-25441313, Professional AI Talent Development Program for Multimodal AI Agents, Contribution: 50\%).
 
This research was supported by Culture, Sports and Tourism R\&D Program through the Korea Creative Content Agency grant funded by the Ministry of Culture, Sports and Tourism in 2024 (Project Name: Development of Co-Pilot technology for automatic completion of generative AI-based 3D Webtoon, Project Number: RS-2024-00400004, Contribution Rate: 50\%).

\section{Generative AI Use Disclosure}
Generative AI tools were used to assist with English translation and language editing of the manuscript. The scientific content, analysis, and conclusions were developed by the authors. All authors take full responsibility for the content of the paper and approve its submission.

\bibliographystyle{IEEEtran}
\bibliography{mybib}

\end{document}